\documentclass{ws-p8-50x6-00}

\begin{document}

\title{Nucleon Magnetic Moments, their Quark Mass Dependence and Lattice QCD 
Extrapolations}

\author{T.R. Hemmert$^A$ and W. Weise$^{A,B}$}

\address{$^A$ Theoretische Physik, Physik Department, TU M{\" u}nchen\\  
D-85747 Garching, Germany}

\address{$^B$ ECT*, Villa Tambosi, I-38050 Villazzano (Trento), Italy}  


\maketitle


\section{Introduction}
The chiral symmetry of QCD is spontaneously broken at low energies, leading 
to the appearance of Goldstone Bosons. For 2-flavor QCD we identify the 
resulting 3 Goldstone Bosons with the 3 physical pion states, the lowest 
lying modes in the hadron spectrum. In addition to being spontaneously 
broken, chiral symmetry is broken explicitly via the non-zero quark mass 
$\hat{m}$ in the QCD lagrangian. This explicit breaking is responsible for 
the non-zero masses $m_\pi$ of the pions. In the now experimentally 
established large-condensate scenario with parameter $B_0$ one obtains
\begin{eqnarray}
m_\pi^2&=&2\,\hat{m}\,B_0\left\{1+{\cal O}(\hat{m}B_0)\right\}
\end{eqnarray}
for this connection. At low energies QCD is represented by a chiral effective 
field theory ($\chi$EFT)
with the dynamics governed by the Goldstone Bosons, coupling to matter 
fields and external sources. The important aspect for this work is the fact 
that this $\chi$EFT incorporates {\em both} the information on the 
spontaneous {\em and} on the explicit breaking of chiral symmetry. We 
report on recent work~\cite{paper}
utilizing $\chi$EFT to study the quark mass (pion mass) dependence of 
the magnetic moments of the nucleon.

\section{The Calculation}
\begin{figure}[t]
\epsfxsize=18pc 
\epsfysize=15pc
\epsfbox{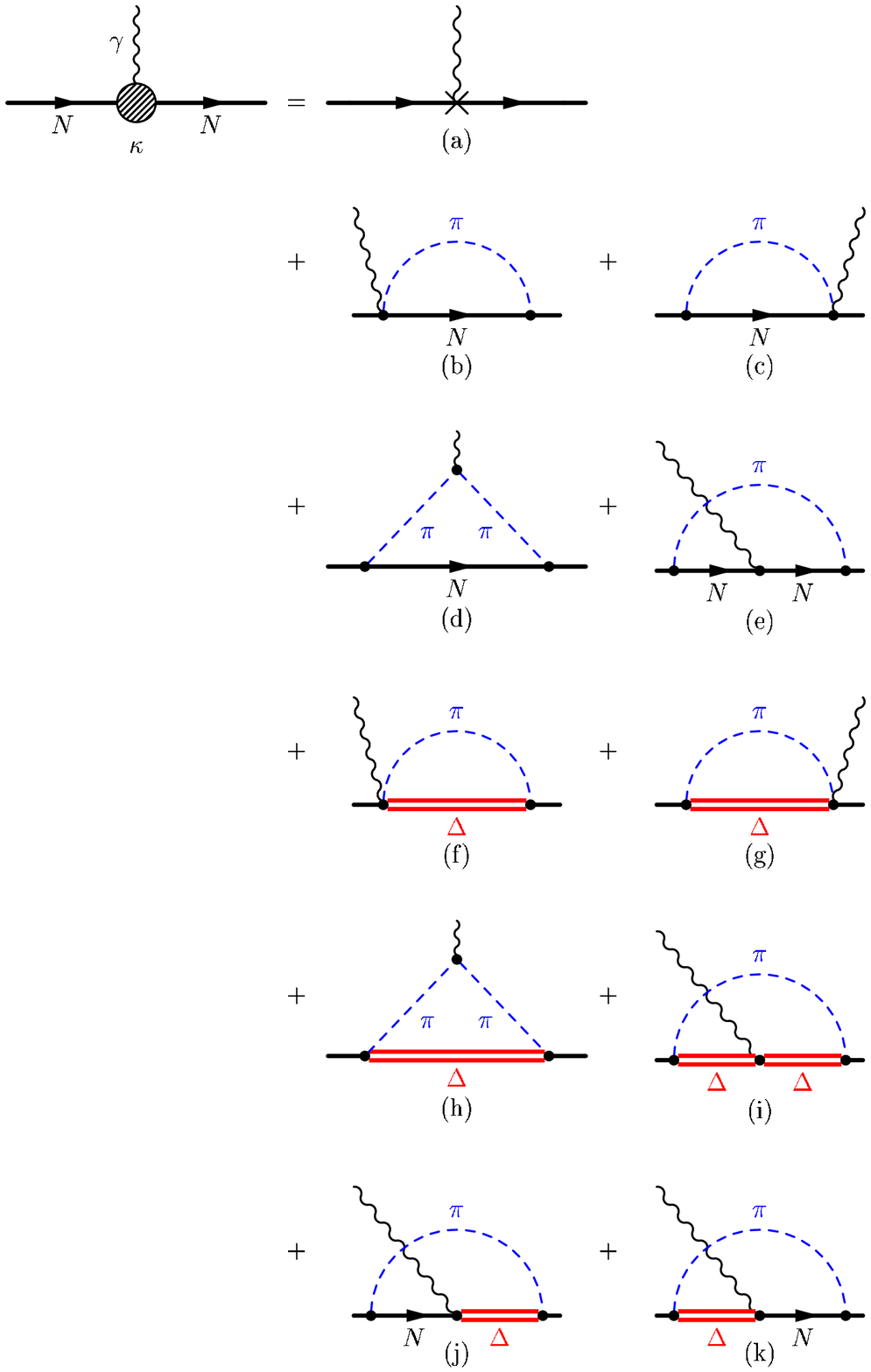} 
\caption{\label{diags}}
\end{figure}

We use $\chi$EFT with pions, nucleons and deltas as explicit degrees of 
freedom. When including matter fields with differing masses---as is the case 
between $\Delta$(1232) and the nucleon---one has to make a decision on the 
power counting one employs. Throughout this work we follow the so called 
Small Scale Expansion (SSE) of refs.\cite{review}. However, in one crucial 
aspect we differ from refs.\cite{review}: For the leading order $N\Delta$ 
transition lagrangian we employ
\begin{eqnarray}
{\cal L}^{(1)}_{N\Delta}&=&\bar{T}^\mu_i\left[c_A\,w^i_\mu+
c_V\,i\,f_{\mu\nu}^{+\,i}S^\nu\right]N_v\,+\,h.c.\; ,
\label{eq:cv}
\end{eqnarray}
which treats vector and axial-vector couplings $c_V,\;c_A$ to this transition 
on a symmetric footing. Usually the (leading order) $N\Delta$ vector coupling 
is relegated to subleading order based on standard (``naive'') power counting 
arguments. Nevertheless, we find it necessary~\cite{paper} to resort to 
Eq.(\ref{eq:cv}) to capture essential quark-mass dependent effects in the 
anomalous magnetic moments already at leading one-loop order, resulting in a 
better behaved perturbative expansion. Our goal is to study the quark (pion) 
mass dependence of the magnetic moments of the nucleon. Treating the 
electromagnetic field as an external vector source, to leading one-loop 
order---according to SSE---one has to evaluate 11 diagrams, displayed in 
Fig.\ref{diags}.

\section{Isovector Anomalous Magnetic Moment}

\begin{figure}[t]
\epsfxsize=20pc 
\epsfysize=12pc
\epsfbox{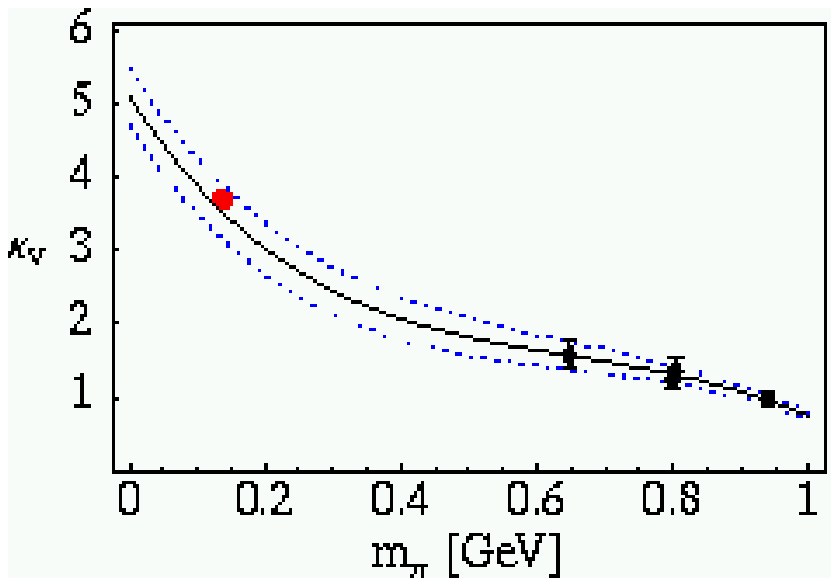} 
\caption{\label{C}}
\end{figure}

For the isovector anomalous magnetic moment one obtains
\begin{eqnarray}
\kappa_v&=&\kappa_{v}^0-\frac{g_A^2\,m_\pi M}{4\pi F_\pi^2}               
  +   \frac{2 c_A^2 \Delta M}{9\pi^2F_\pi^2}
                              \left\{\sqrt{1-\frac{m_\pi^2}{\Delta^2}}\log
  \left[R(m_\pi)\right]+\log\left[\frac{m_\pi}{2\Delta}\right]
                              \right\} \nonumber\\
          & &               -   8 E_1(\lambda) Mm_\pi^2
                         +  \frac{4c_A c_V g_A M m_\pi^2}{9\pi^2F_\pi^2}
\log\left[\frac{2\Delta}{\lambda}\right] 
+  \frac{4c_A c_V g_A M m_\pi^3}{27\pi F_\pi^2\Delta}
                            \nonumber\\
          & &            -   \frac{8 c_A c_V g_A \Delta^2 M}{27\pi^2F_\pi^2}
                              \left\{\left(1-\frac{m_\pi^2}{\Delta^2}
\right)^{3/2}\log\left[R(m_\pi)\right]+\left(1-\frac{3m_\pi^2}{2\Delta^2}
\right)
                              \log\left[\frac{m_\pi}{2\Delta}\right]
                              \right\} \nonumber\\
          & &            +N^2LO \;,\label{sc}
\end{eqnarray}
where $\Delta$ is the nucleon-delta mass difference. Most of the parameters 
in this expression are known and specified in ref.~\cite{paper}, except for 
$\kappa_v^0,\,c_V,\,E_1$. At a chosen regularization scale $\lambda$ we fit 
these 3 parameters to reproduce quenched lattice results for $\kappa_v$ 
reported in ref.~\cite{adelaide}. Note that these lattice data correspond 
to lattice pions heavier than 600 MeV. With the parameters now fixed one 
obtains the full curve in Fig.\ref{C}, which at $m_\pi=140$ MeV comes very 
close to the physical isovector anomalous magnetic moment, indicated by the 
full circle. A priori it is not guaranteed that this extrapolation curve over 
such a wide range of quark (pion) masses would come anywhere near to the 
physical value, but remarkably it does so, albeit with a large error band 
(dashed curves). We also note that our approach rests on the assumption that 
for lattice data with effective pion masses larger than 600 MeV the 
differences between quenched and fully dynamical lattice simulations are 
small~\cite{paper}, allowing us to utilize ``Standard'' instead of 
``Quenched'' $\chi$EFT methods.

\section{Isoscalar Anomalous Magnetic Moment}
To the same leading-one loop order in SSE one only obtains analytic quark 
mass dependence for the isoscalar anomalous magnetic moment $\kappa_s$:
\begin{eqnarray}
\kappa_s&=&\kappa_{s}^0-8\,E_2\,M\,m_\pi^2+N^2LO\;.\label{eq:kappascalar}
\end{eqnarray}
The 2 unknown couplings $\kappa_s^0,\,E_2$---parameterizing short-distance 
physics beyond the realm of $\chi$EFT---can again be fitted to lattice 
data~\cite{paper}.

\section{Comparison with Pade-Formula}
\begin{figure}[t]
\epsfxsize=20pc 
\epsfysize=12pc
\epsfbox{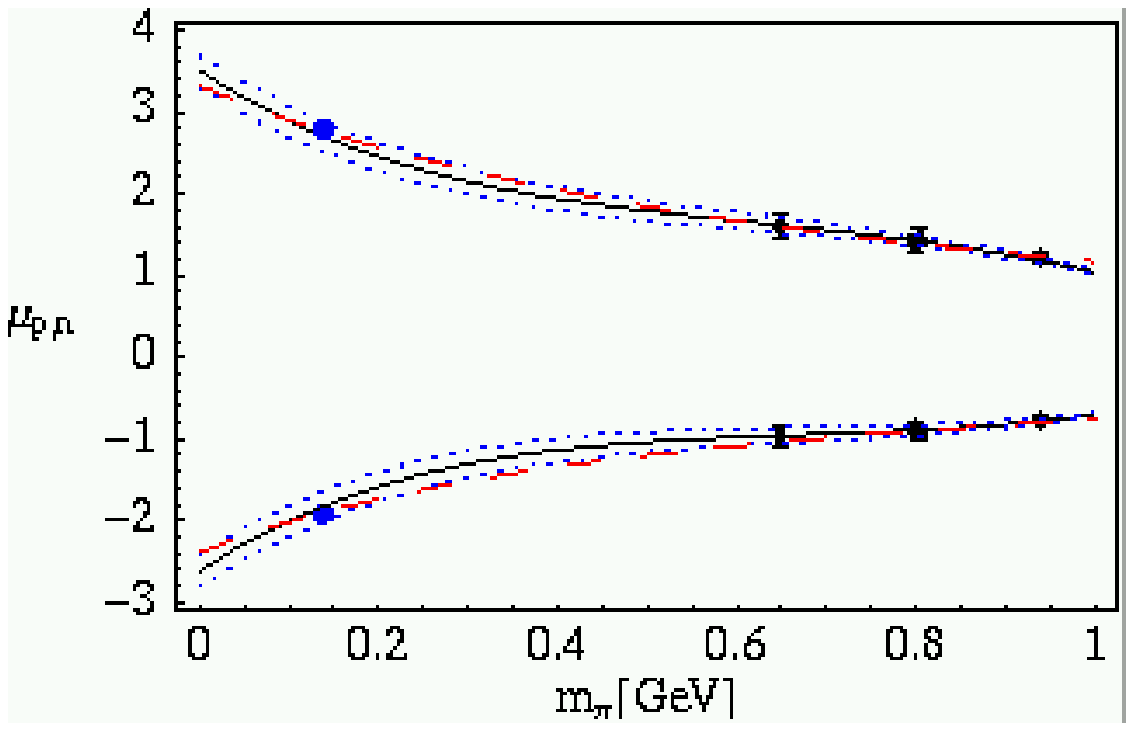} 
\caption{\label{E}}
\end{figure}

Combining isovector and isoscalar results one obtains the quark (pion) mass 
dependence of the magnetic moments of proton and neutron, as shown in the 
full line of Fig.\ref{E}. Surprisingly our result is rather close---at least 
within the present error band---to the Pade-fit extrapolation formula of the 
Adelaide group~\cite{adelaide}, shown as the dashed curve. 

\section*{Acknowledgments}
The authors acknowledge partial support by BMBF and DFG.

\end{document}